\documentclass[aps,prl,reprint,twocolumns]{revtex4-1}

\usepackage{graphicx,xcolor,soul}

\newcommand{\Schw}{Schwarzschild}

\newcommand{\beq}{\begin{equation}}
\newcommand{\eeq}{\end{equation}}
\newcommand{\bea}{\begin{eqnarray}}
\newcommand{\eea}{\end{eqnarray}}

\def\d{\dif}

\def\RS{\Sigma}

\providecommand{\dif}{\mathrm{d}} \def\d{\dif}

\def\mir{\mathrm{r}}

\def\mip{\mathrm{\phi}}

\begin{document}

\title{Epicyclic oscillations in spinning particle motion around Kerr black hole  applied in models fitting the quasi-periodic oscillations observed in quasars and microquasars}

\author{
Misbah Shahzadi and Martin Kolo\v{s} and Zden\v{e}k Stuchl{\'i}k and Yousaf Habib
}

\address{$^{1}$Department of Mathematics, COMSATS University Islamabad, Lahore Campus-54000, Pakistan.}
\address{$^{2}$Research Centre for Theoretical Physics and Astrophysics, Institute of Physics, Silesian University in Opava, Bezru{\v c}ovo n{\'a}m.13, CZ-74601 Opava, Czech Republic.}
\address{\normalfont {misbahshahzadi51@gmail.com, martin.kolos@physics.slu.cz, zdenek.stuchlik@physics.slu.cz, yhabib@cuilahore.edu.pk}}

\begin{abstract}
The study of the quasi-periodic oscillations (QPOs) of X-ray flux observed in the stellar-mass black hole (BH) binaries or quasars can provide a powerful tool for testing the phenomena occurring in strong gravity regime.
We thus fit the data of QPOs observed in the well known quasars as well as microquasars in the framework of the model of geodesic oscillations of Keplerian disks modified for the epicyclic oscillations of spinning test particles orbiting Kerr BHs. We show that the modified geodesic models of QPOs can explain the observational fixed data from the quasars and microquasars but not for all sources. We perform a successful fitting of the high frequency QPOs models of epicyclic resonance and its variants, relativistic precession and its variants, tidal disruption, as well as warped disc models, and discuss the corresponding constraints of parameters of the model, which are the spin $S$ of the test particle, mass $M$ and spin $a$ of the BH.
\end{abstract}

\maketitle

\section{Introduction} \label{intro}

The motion of a spinning test-body in a fixed gravitational background is a long-standing problem in general relativity \cite{Prap:1951:proRSPSA:}. The rotation of a system plays a very important role in astrophysics. The spin or angular momentum can completely alter the evolution of the system. In a dynamical system, the spin or rotation is quite important as it may cause the chaotic behavior. It is known that the spin-orbit interaction produces the chaos in Newtonian gravity.
It might be true in some relativistic system such as the evolution of a binary system. The motion of coalescing binary systems of neutron stars and BHs is quite interesting to explore because they are promising sources of gravitational waves.

The Quasi-periodic oscillations in X-ray flux light curves have long been observed in stellar-mass BH binaries and considered as one of the most efficient tests of strong gravity models. These variations appear very close to the BH, and present the frequencies that scale inversely with the mass of the BH. The current technical possibilities to measure the frequencies of QPOs with high precision allow us to get useful knowledge about the central object and its background. According to the observed frequencies of QPOs, which cover the range from few mHz up to $0.5$kHz, different types of QPOs were distinguished. Mainly, these are the high frequency (HF) and low frequency (LF) QPOs with frequencies up to 500~Hz and up to 30~Hz, respectively. The oscillations of HF QPOs in BH microquasars are usually stable and detected with the twin peaks which have frequency ratio close to $3:2$ \cite{Rem-McCli:2006:ARAA:}. However, this phenomenon is not universal, the HF QPOs have been observed in only $11$ out of $7000$ observations of $22$ stellar mass BHs \cite{Bel-et-al:2012:MNRAS}. The oscillations usually occur only in specific states of hardness and luminosity, moreover, in X-ray binaries, HF QPOs occur in “anomalous” high-soft state or steep power law state, both corresponding to a luminous state with a soft X-ray spectrum.

Microquasars are binary systems composed of a BH and a companion (donor) star; matter floating from the companion star onto the BH forms an accretion disk and relativistic jets - bipolar outflow of matter along the BH - accretion disk rotation axis. Due to friction, matter of the accretion disk becomes to be hot and emits electromagnetic radiation, also in X-rays in vicinity of the BH horizon.

Applying the methods of spectroscopy (frequency distribution of photons) and timing (photon number time dependence) for particular microquasars, one can extract a useful information regarding the range of parameters of the system \cite{Rem-McCli:2006:ARAA:}. In this connection, the binary systems containing BHs, being compared to neutron star systems, seem to be promising due to the reason that any astrophysical BH is thought to be a Kerr BH (corresponding to the unique solution of general relativity in 4D for uncharged BHs which does not violate the no hair theorem and the weak cosmic censorship conjecture) that is determined by only two parameters: the BH mass $M$ and the dimensionless spin $|a|\leq~1$.

After the first detection of QPOs, there were various attempts to fit the observed QPOs, and different models have been proposed, such as the disko-seismic models, hot-spot models, warped disk model and many versions of resonance models. The most extended are thus the so called geodesic oscillatory models where the observed frequencies are related to the frequencies of the geodesic orbital and epicyclic motion - for review see \cite{Stu-Kot-Tor:2013:ASTRA:}.
It is particularly interesting that the characteristic frequencies of HF QPOs are close to the values of the frequencies of test particle, geodesic epicyclic oscillations in the regions near the innermost stable circular orbit (ISCO) which makes it reasonable to construct the model involving the frequencies of oscillations associated with the orbital motion around Kerr BHs \cite{Stu-Kot-Tor:2013:ASTRA:}. However, until now, the exact physical mechanism of the generation of HF QPOs is not known, since none of the models can fit the observational data from different sources \cite{Bur:2005:RAG:}. Even more serious situation has been exposed in the case of HF QPOs related to accretion disks orbiting supermassive  BHs in active galactic nuclei \cite{Kot-etal:2020:AAP:,Smi-Tan-Wag:2021:APJ:}. One possible way to overcome this issue is related to the electromagnetic interactions of slightly charged matter orbiting a magnetized BH \cite{Kol-Stu-Tur:2015:CLAQG:,Kol-Tur-Stu:2017:EPJC:,Stu-etal:2020:Universe:}. Here, we concentrate on different possibility related to the internal rotation of accreating matter.

In the present paper, we consider the orbital and epicyclic motion of neutral spinning test particles orbiting a Kerr BH. We look especially for the existence and properties of the harmonic or QPOs of neutral spinning test particle in the background of Kerr BH. The quasi-harmonic oscillations around a stable equilibrium location and the frequencies of these oscillations are then compared with the frequencies of the HF and LF QPOs observed in microquasars GRS 1915+105, GRO 1655-40, XTE 1550-564, and XTE J1650-500  \cite{McC-etal:2011:CLAQG:,Tor-etal:2005:ASTRA:} and quasars TON~S~180, ESO~113-G010, 1H0419-577, RXJ~0437.4-4711, 1H0707-495, RE~J1034+396, Mrk~766, ASASSN-14li, MCG-06-30-15, XMMU~J134736.6+173403, Sw~J164449.3+573451, MS~2254.9-3712 \cite{Smi-Tan-Wag:2021:APJ:}.

Throughout the present paper, we use the spacelike signature $(-,+,+,+)$. Greek indices are taken to run from $0$ to $3$. However, for expressions having astrophysical relevance we use the physical constants explicitly.

\begin{figure*}
\includegraphics[width=\hsize]{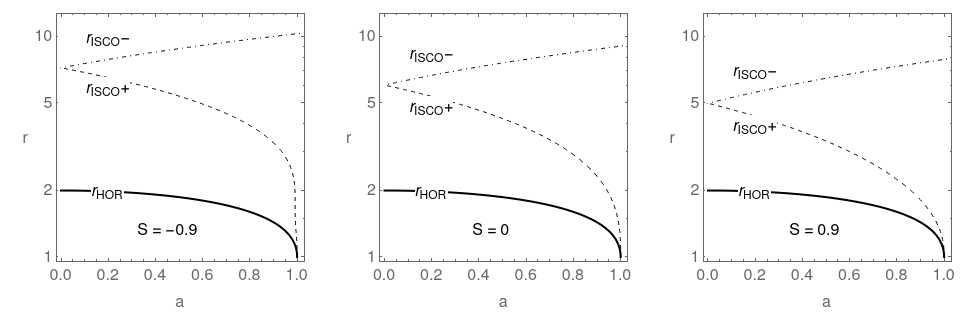}
\includegraphics[width=\hsize]{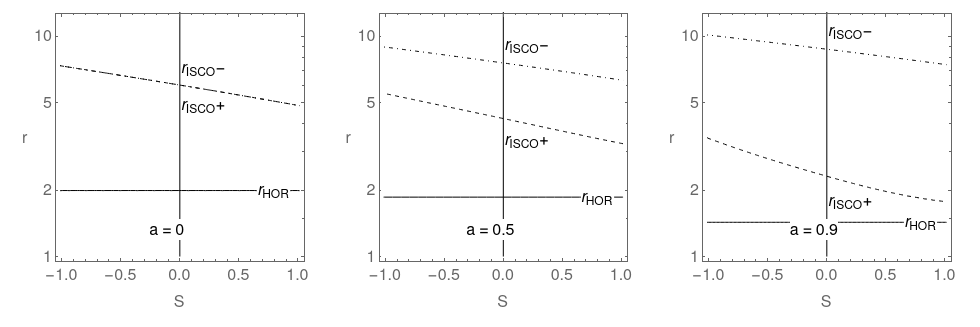}
\caption{The position of horizon and ISCO of a spinning neutral test particle orbiting Kerr BH. The first row is plotted for different values of spin parameter $S$ while second row is plotted for different values of rotation parameter $a$.
\label{ISCO}
}
\end{figure*}

\section{Spinning particle dynamics} \label{MP-EQs}

The motion of a spinning test particle with mass $\mu$, and spin $S$ can be characterized by the Mathisson-Papapetrou (MP) equations and their formulation reads \cite{Prap:1951:proRSPSA:,Dixon:1970:RSPSA:,Dixon:1974:RSPTA:}
\bea \label{spin1}
\frac{dx^\alpha}{d\tau} &=& v^\alpha,\\\label{spin2}
\frac{D p^\alpha}{D\tau} &=& -\frac{1}{2} R^{\alpha}_{\mu\nu\rho} S^{\nu\rho}v^\mu,\\\label{spin3}
\frac{D S^{\alpha\beta}}{D\tau} &=& p^\alpha v^\beta - v^\beta p^\alpha,
\eea
where $\tau$ is the proper time, $D/D\tau$ is the covariant derivative along the particle trajectory, $R^{\alpha}_{\mu\nu\rho}$ defines the Riemann curvature tensor, $S^{\alpha\beta}$ is the antisymmetric spin tensor, $v^\alpha$ (tangent to the particle's worldline), and $p^\alpha$ represents the four-velocity and four-momentum of a test particle, respectively.

The right-hand side of Eq. (\ref{spin2}) indicates the spin-orbit coupling through a strong gravitational field. The spinning particle does not follow the geodesic path. In this situation, the four-momentum deviates from the geodesic trajectory (non-spinning case) because of the spin-curvature force which produces due to the coupling of spin tensor with the Riemann tensor. In the case of vanishing spin ($S^{\alpha \beta}=0$) or flat spacetime ($R^{\alpha}_{\mu\nu\rho}=0$), we recover the geodesic equation $D p^\alpha / D\tau = 0$.

The MP equations are the first order non-linear ordinary differential equations but not a closed set, i.e., in order to evolve the system, there are less equations than necessary. Thus one needs further conditions to specify a reference point about which the spin and momentum of the particle can be calculated. This reference point can be taken as a centre of mass of the body. Different spin supplementary conditions (SSCs) have been proposed to close the set of MP equations \cite{Sem:MNRAS:1999:,Kyr-Sem:2007:MNRAS:}. Since the SSC fixes the center of the mass, and different SSCs define different centers, for each SSC we have a different world line, hence, each SSC specifies a different evolution of the MP equations. Thus the equations of motion of a spinning particle are not unique. In order to evolve the MP equations, we use the Tulczyjew-Dixon (TD) SSC defined by \cite{Dixon:1970:RSPSA:,Dixon:1974:RSPTA:}
\beq \label{spin-cond}
p_{\sigma} S^{\sigma\rho} =0.
\eeq
Using the above condition, one can write the explicit relation between $v^\mu$ and $p^\mu$ as \cite{Ehl-Rud:1977:GReGr:}
\beq
v^\mu = N \left( p^{\mu} + \frac{2 S^{\mu\nu} p^{\lambda}  R_{\nu\lambda\rho\sigma}  S^{\rho\sigma}}{4 \mu^{2} + S^{\alpha\beta} R_{\alpha\beta\gamma\delta} S^{\gamma\delta} }\right),
\eeq
where the normalization constant $N$ can be fixed using the constraint $v_{\alpha}v^{\alpha}=-1$. The tensor and vector formulations of the spin for TD SSC are related by
\beq
S^{\alpha\beta}= -\eta^{\alpha\beta\gamma\delta} S_{\gamma} u_{\delta},
\eeq
and
\beq
S_\alpha = -\frac{1}{2}\eta_{\alpha\beta\mu\nu} u^{\beta} S^{\mu\nu},
\eeq
where $\eta_{\alpha\beta\mu\nu} = \sqrt{-g} ~ \epsilon_{\alpha\beta\mu\nu}$ is the Levi-Civita tensor, $\epsilon_{\alpha\beta\mu\nu}$ denotes the Levi-Civita symbol, $u^{\alpha}=p^{\alpha}/\mu$ ($=p^{\alpha}$ in normalized units) represents a unit vector parallel to the momentum. One can prove that the mass $\mu$ and spin magnitude $S$ of the particle defined by
\bea \label{Int-motion1}
\mu &=& \sqrt{- p^\alpha p_{\alpha}},\\\label{Int-motion2}
S^2 &=& S^\alpha S_\alpha = \frac{1}{2} S^{\alpha\beta} S_{\alpha\beta},
\eea
are the constant of motion independent from the symmetry of the background spacetime. However, the constant of motion dependent on the symmetry of the background spacetime can be constructed using the Killing vector $\xi$ as
\beq \label{killing-field1}
C({\xi}) = p^{\alpha} \xi_{\alpha} -\frac{1}{2} S^{\alpha\beta} \xi_{\alpha ; \beta}.
\eeq
The numerical study of MP equations entails a bunch of interesting numerical challenges. The efficient integration of equations of motion over a long time interval needs the structure preserving algorithms \cite{Hairer:2006:}, i.e., symplectic schemes, which have been successfully applied for simulations in different fields of general relativity. Furthermore, the MP equations have no Hamiltonian structure, therefore one would expect usual symplectic integration schemes to lose their theoretical advantage over ordinary, not so efficient ones.

\begin{figure*}
\includegraphics[width=\hsize]{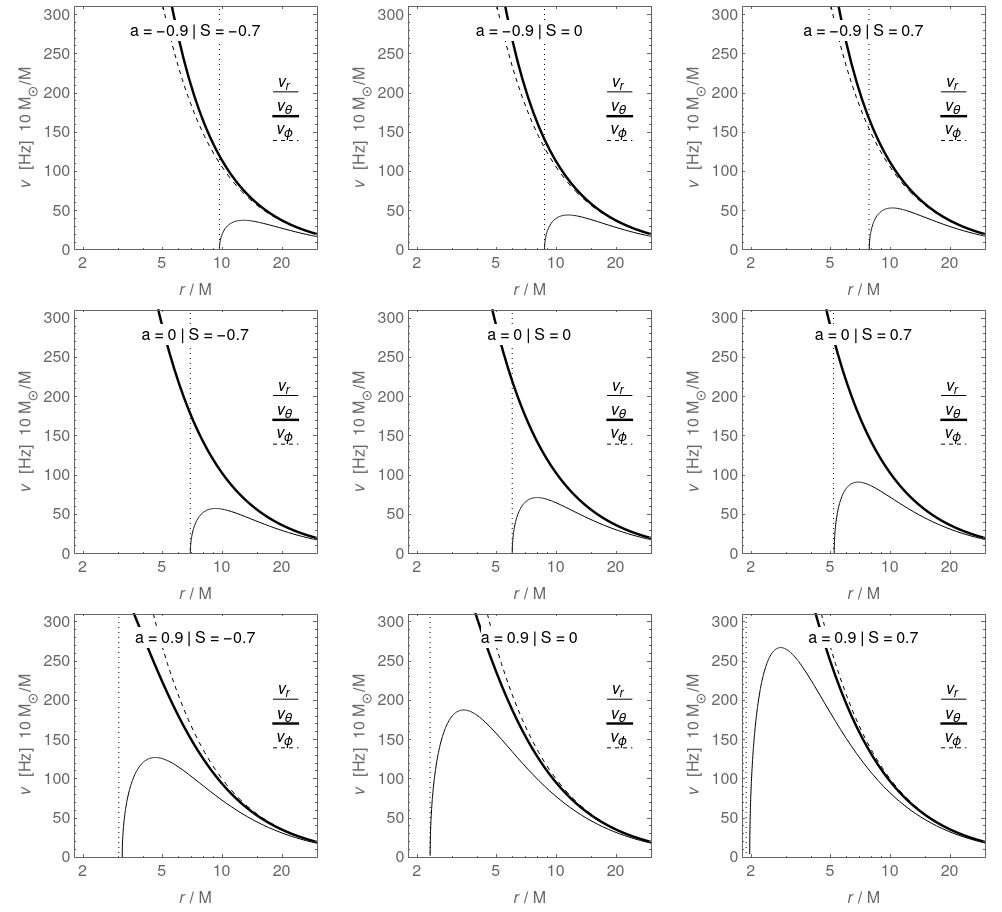}
\caption{Radial profiles of frequencies of small harmonic oscillations $\nu_{r}$, $\nu_{\theta}$ and $\nu_{\phi}$ of neutral spinning test particles in the background of Kerr BH having mass $M=10 M_{\odot}$ measured by a static distant observer. The dotted lines represent the ISCO position. \label{frequencies} }
\end{figure*}

\subsection{Rotating Kerr black hole} \label{secKBH}

The geometry of the Kerr BH can be described by the line element
\beq
\d s^2 = g_{tt}\d{t}^2 + g_{rr}\d{r}^2 + g_{\theta\theta}\d\theta^2 + g_{\phi\phi}\d\phi^2 +2g_{t\phi}\d{t}\d\phi, \label{Kerr-MOGMetric}
\eeq
with the nonzero components of the metric tensor $g_{\mu\nu}$ taking in the standard Boyer-Lindquist coordinates the form
\bea
g_{tt} &=& -\left(\frac{\Delta-a^{2}\sin^{2}\theta}{\RS}\right), \quad
g_{rr} = \frac{\RS}{\Delta}, \quad g_{\theta\theta} = \RS,\nonumber\\
g_{\phi\phi} &=& \frac{\sin^{2}\theta}{\RS}\left[(r^{2}+a^{2})^{2}-\Delta
a^{2}\sin^{2}\theta\right], \nonumber\\
g_{t\phi}&=&\frac{a\sin^{2}\theta}{\RS}\left[\Delta-(r^{2}+a^{2})\right] ,
\label{MetricCoef}
\eea
where
\bea
\Delta &=& r^2 - 2GMr + a^2, \nonumber\\ \RS &=& r^2 + a^2 \cos^2\theta.
\eea
The outer horizon is situated at
\beq
r_{+} = M + \sqrt{ M^2 -a^2 }. \label{horizon}
\eeq

The geometrical structure of horizon of Kerr BH for different values of $a$ is shown in Fig. {\ref{ISCO}}. We see that the horizon is only dependent on the rotation parameter $a$. It is observed that radius of horizon decreases when BH rotates rapidly.

The circular orbits are very important from the astrophysical point of view as they govern the thin (Keplerian) accretion disks, and even the toroidal fluid configurations. The position of the smallest stable circular orbit so called ISCO is illustrated in Fig.~\ref{ISCO}. The co-rotating particles of ISCO shift towards the BH horizon with increase of the BH rotation $a$, while counter-rotating particles shift away from the BH horizon. However, when spin of particle is parallel to that of the BH, then for increasing values of spin parameter $S$, both counter-rotating as well as co-rotating particles of ISCO shift towards the BH horizon, while for antiparallel case, both counter and co-rotating particles shift away from the BH. It is noted that the spinning particle has smaller (greater) ISCO as compared to the non-spinning particle. It is noted that the non-spinning particle has greater (smaller) ISCO as compared to spinning case when spin of particle is parallel to that of the BH (antiparallel case).

\begin{figure*}
\includegraphics[width=\hsize]{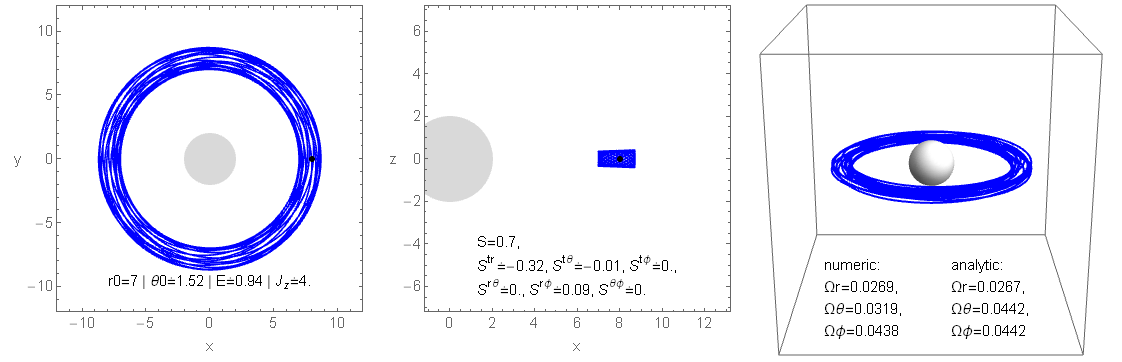}
\includegraphics[width=\hsize]{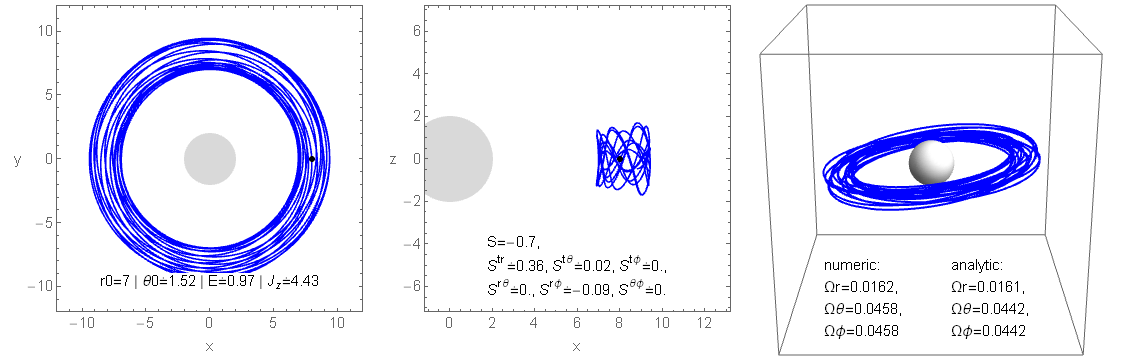}
\caption{
Perturbed circular orbit for spinning particles around \Schw{} BH; numerical frequencies (from trajectory Fourier spectra) and analytical frequencies (Eqs. \ref{3}-\ref{4}) are given. Numerical and analytical frequencies are identical for exactly circular orbit only, but as we can see they are quite similar even for large deviation for circular orbit. The initial conditions used for components of spin-tensor $S^{\alpha\beta}_0$ for each trajectory are given in the middle plot.
\label{trajectories} }
\end{figure*}

\subsection{Spinning particle fundamental frequencies} \label{oscillations}

In order to study the oscillatory motion of neutral particle one can use the perturbation of equations of motion around the stable circular orbits. If a test particle is slightly displaced from the equilibrium state corresponding to a stable circular orbit situated in an equatorial plane, the particle will start to oscillate around the stable orbit realizing thus epicyclic motion governed by linear harmonic oscillations. If the frequencies of small harmonic oscillations measured by distant observer are expressed in physical units, one need to extend the corresponding dimensionless form by the factor $c^3/GM$. Thus the frequencies of neutral spinning particle orbiting Kerr BH measured by distant observer are given by
\beq\label{nu_rel}
\nu_{j}=\frac{1}{2\pi}\frac{c^{3}}{GM} \, \Omega_{j}[{\rm Hz}],
\eeq
where $j\in\{r,\theta,\phi\}$. The dimensionless radial $(\Omega_{r})$, latitudinal ($\Omega_{\theta}$) and axial ($\Omega_{\phi}$) angular frequencies measured by a distant observer for a neutral spinning test particle harmonic oscillations around a Kerr BH are given by \cite{Tan-et-al:2013:PhRvD}
\bea \label{3}
\Omega_{r}^{2}&=& \frac{r^{3/2}(\alpha - 3 a^2\pm 8 a \sqrt{r}) + \beta \delta }{r^{7/2} \left(a+r^{3/2}\right)^2} + O(S^2), \\\label{3a}
\Omega_{\theta}^{2}&=& \frac{r \left(\left(3 a^2\mp 4 a \sqrt{r}\right)+r^2\right)^3}{\left(a+r^{3/2}\right)^2 + (X+Y)^2 } + O(S^2), \\\label{4}
\Omega_{\phi}^{2}& = & \frac{1}{\left(a+r^{\frac{3}{2}}\right)^2},
\eea
\bea
\alpha &=& r (r-6), \quad \beta = (r-3) \sqrt{r}\pm 2 a,\\\
\delta &=& 6 S \left(\pm \sqrt{r} - a\right), \quad X = r^{7/2}+9 a \sqrt{r} S, \\\
Y &=& r^{3/2} \left(3 a^2\mp 4 a \sqrt{r}\right) \mp 6 a^2 S  -3 a r^{3/2} S,
\eea
where the upper and lower signs correspond to the co-rotating and counter rotating orbits, respectively. The radial profiles of the frequencies $\nu_{j}$ of small harmonic oscillations of neutral particle measured by a distant static observer are shown in Fig. {\ref{frequencies}}, for different values of spin $S$ and rotation parameter $a$. In the case of non-rotating BHs (Schwarzschild), radial ($\nu_{r}$) and latitudinal ($\nu_{\theta}$) frequencies coincide, but for rotating BHs (Kerr), $\nu_{r}$ and $\nu_{\theta}$ can be observed with different profiles. The profiles of frequencies lowering down or up depending on the direction of the spin $S$. The presence of spin $S$ contributes to lowering down the peaks of all the frequencies $\nu_{r}$, $\nu_{\theta}$ and $\nu_{\phi}$ when particle has antiparallel spin to that of the BH, while the peaks rises up when spin of the particle is parallel to that of the BH. It is also noted that radial profiles of frequencies shift towards the BH with the increase of the rotation  parameter $a$ of the BH. In the case of non-spinning particle, the peaks of the radial profiles of frequencies are lower (high) as compared to the case of the spinning particle when spin rotation is parallel to that of BH (antiparallel to that of BH).

The newly introduced particle spin parameter $S$ allow the greatest diversity in radial profiles of fundamental frequencies as shown in Fig.~\ref{frequencies}. The spin effect on particle frequencies is very relevant in the region close ($\nu_r=0$) or below the ISCO ($r<r_{\rm ISCO}$) position as well as for high spin values $S,a\rightarrow~1$. For $r<r_{\rm ISCO}$, where the oscillations are unstable against the radial perturbations, one can assume the behavior fundamentally different from the non-spinning case ($S=0$) especially when nonlinear terms in spin parameter $S$ will be included. In this study we are interested in stable particle oscillations above ISCO, where the fundamental frequencies are similar to non-spinning case but slightly shifted.

The behavior of the particle fundamental frequencies lowering down or increasing up depends on the direction of the BH rotation parameter $a$ and test particle spin $S$. Both parameters $a$ and $S$ contribute to increase the fundamental frequencies, however, higher frequencies can be observed when the spin is aligned to the z-axis ($S>0$) while lower frequencies in the opposite case ($S<0$). It is also noticed that co-rotating particles ($a>0$) have higher frequencies as compared to contra-rotating particles ($a<0$).

\subsection{Numerical Analysis}

We numerically integrate the equations of motion of a neutral spinning test body in the background of rotating Kerr BH using the Gauss Runge Kutta method of order four and investigate the fundamental frequencies numerically with the help of the Fourier transformation. We plotted perturbation of circular orbit and calculate the corresponding fundamental frequencies, see Fig.~\ref{trajectories}. In order to find out appropriate initial conditions, we use spin-1 form instead of spin tensor. We can not choose the initial conditions randomly to evolve the system, instead, our initial data should satisfy the following constraint equations
\bea \label{data1}
E &=& -p_t - \frac{1}{2} g_{t\alpha, \beta} \eta^{\alpha\beta\rho\delta} S_{\rho} p_{\delta},\\\label{data2}
J_z &=& p_\phi + \frac{1}{2} g_{\phi \alpha, \beta} \eta^{\alpha\beta\rho\delta} S_{\rho} p_{\delta},\\\label{data3}
0 &=& g^{\alpha\beta} S_\alpha p_\beta. \\\label{data4}
\mu^2 &=& -g^{\alpha\beta} p_\alpha p_{\beta}, \\\label{data5}
S^2 &=& g^{\alpha\beta} S_\alpha S_\beta.
\eea
Initially we consider the motion of a spinning test particle on an equatorial plane, and assume that the spin-vector of the particle is aligned with the orbital angular momentum, i.e., the direction of spin is parallel to the rotational axis. Thus we have
\beq
\theta = \frac{\pi}{2}, \quad p^{\theta} = 0, \quad S^{\alpha} = S^{\theta}\delta^{\alpha}_{\theta}.
\eeq
Under these assumptions, using the Eq. (\ref{data5}), the spin-vector can be expressed in terms of spin magnitude $S$ as
\beq
S_{\theta} = -\sqrt{g_{\theta \theta}} ~ S,
\eeq
with $S > 0 ~ (S < 0)$ corresponding to a spin-vector (anti-)aligned with the orbital angular momentum, which by convention is always pointing along the positive z-direction.

For the existence of circular orbits, we set $p^{r} = 0$, the energy $E$ and angular momentum $J_{z}$ can be written as \cite{Tan-et-al:2013:PhRvD}
\bea
E &=& \frac{\sqrt{\Delta}}{r} p^{t} + \frac{1}{r} \left(a + \frac{M S}{\mu} \right)p^{\phi},\\
 J_{z} &=& \frac{\sqrt{\Delta}}{r} \left(a + \frac{S}{\mu}\right) p^{t} \nonumber \\
 && + \frac{1}{r} \left(a^{2} + r^{2} + \frac{a S}{\mu} \left(1 + \frac{M}{r} \right)\right)p^{\phi}.
\eea
In the present situation, we have only two non-zero components of spin-tensor given by
\bea
S^{tr} = - S^{\theta}p_{\phi} = \frac{p_{\phi}}{r} S, \quad S^{r\phi} = -S^{\theta} p_{t} = \frac{p_{t}}{r}S.
\eea
For our convenience, we introduce the dimensionless quantities as
\beq
r=\frac{r}{M}, \quad a=\frac{a}{M}, \quad S=\frac{S}{\mu M}, \quad J_{z}=\frac{J_z}{\mu M}, \quad E=\frac{E}{\mu}.
\eeq
The dimensionless quantities are equivalent to the dimensional quantities when one sets $\mu = M =1$.

The spinning particle fundamental frequencies calculated as perturbation of circular orbit can also be used for oscillations with relatively large amplitudes, see Fig.~\ref{trajectories}. It is shown that the numerically calculated frequencies using Eqs. (\ref{spin1})-(\ref{spin3}) are approximately the same as calculated analytically with Eqs. (\ref{3})-(\ref{4}).

\begin{table*}[!ht]
\begin{center}
\begin{tabular}{c l l l l l}
\hline
Name	&	BH Spin	&	$M_\mathrm{BH}$ [$M_\odot$]	&	$f_\mathrm{QPO}$ [Hz]	&	&	Object Type \\
\hline
XTE J1550-564	&	$0.75 < a < 0.77^{r}$ 	&	$9.1^{+0.6}_{-0.6}$		&	184 	& & microquasar \\
				&	$0.29 < a < 0.62^{r,c}$ &							&	184 	\\
\hline
XTE J1650-500	&	$0.78 < a < 0.8^{r}$	&	$5.0^{+2.0}_{-2.0}$ 	&	250		& & microquasar \\
\hline
GROJ1655-40	& $0.65 < a < 0.75^{c}$ & $5.9^{+0.8}_{-0.8}$ &	300	& & microquasar \\
			& $0.9 < a < 0.998^{r}$	&					  & 300		\\
			& $0.97 < a < 0.99^{r}$	& 					  & 300	\\
\hline
GRS1915+105	& $0.97 < a < 0.99^{r,c}$ &	$9.5-14.4$ &	41	& & microquasar \\
			& 						  &			   &	67	\\
			&						  &			   &	166 \\
\hline
 	&		&	log $M_\mathrm{BH}$	&		&	QPO Band	&	 \\
\hline
TON S 180	&	$ < 0.4$	&	$6.85^{+0.5}_{-0.5}$	&	$5.56\times10^{-6}$	&	EUV	&	NLS1 \\\\
ESO 113-G010	&	0.998	&	$6.85^{+0.15}_{-0.24}$	&	$1.24\times10^{-4}$	&	X	&	NLS1 \\\\
ESO 113-G010	&	0.998	&	$6.85^{+0.15}_{-0.24}$	&	$6.79\times10^{-5}$	&	X	& NLS1 \\\\
1H0419-577	&	$ > 0.98$	&	$8.11^{+0.50}_{-0.50}$	&	$2.0\times10^{-6}$	&	EUV	&	Sy1 \\\\
RXJ 0437.4-4711	&	--	&	$7.77^{+0.5}_{-0.5}$	&	$1.27\times10^{-5}$	&	EUV	&	Sy1 \\\\
1H0707-495	&	$>0.976$	&	$6.36^{+0.24}_{-0.06}$	&	$2.6\times10^{-4}$	&	X	&	NLS1	\\\\
RE J1034+396	&	0.998	&	$6.0^{+1.0}_{-3.49}$	&	$2.7\times10^{-4}$	&	X	&	NLS1	\\\\
Mrk 766	&	$>0.92$	&	$6.82^{+0.05}_{-0.06}$	&	$1.55\times10^{-4}$	&	X	&	NLS1	\\\\
ASASSN-14li	&	$>0.7$	&	$6.23^{+0.35}_{-0.35}$	&	$7.7\times10^{-3}$	&	X	&	TDE	\\\\
MCG-06-30-15	&	$>0.917$	&	$6.20^{+0.09}_{-0.12}$	&	$2.73\times10^{-4}$	&	X	&	NLS1	\\\\
XMMU J134736.6+173403	&	--	&	$6.99^{+0.46}_{-0.20}$	&	$1.16\times10^{-5}$	&	X		\\\\
Sw J164449.3+573451	&	--	&	$7.0^{+0.30}_{-0.35}$	&	$5.01\times10^{-3}$	&	X	&	TDE	\\\\
MS 2254.9-3712	&	--	&	$6.6^{+0.39}_{-0.60}$	&	$1.5\times10^{-4}$	&	X	&	NLS1 \\
\hline
\end{tabular}
\caption{
Observational data for QPOs around stellarmass and supermassive BHs  \cite{Smi-Tan-Wag:2021:APJ:}. In objects where multiple HF QPOs are reported in small integer ratios, only the lowest frequency is given. Superscripts on the spin ranges indicate the measurement methods: Fe~K$\alpha$ reflection spectroscopy ($r$) or continuum fitting ($c$). Restrictions on mass and spin of the BHs located in them, based on measurements independent of the HF QPO measurements given by the optical measurement for mass estimates and by the spectral continuum fitting for spin estimates \cite{Sha-etal:2006:ApJ:,Rem-McCli:2006:ARAA:}.
\label{tab1}
}
\end{center}
\end{table*}

\begin{figure*}
\begin{center}
\includegraphics[width=0.9\hsize]{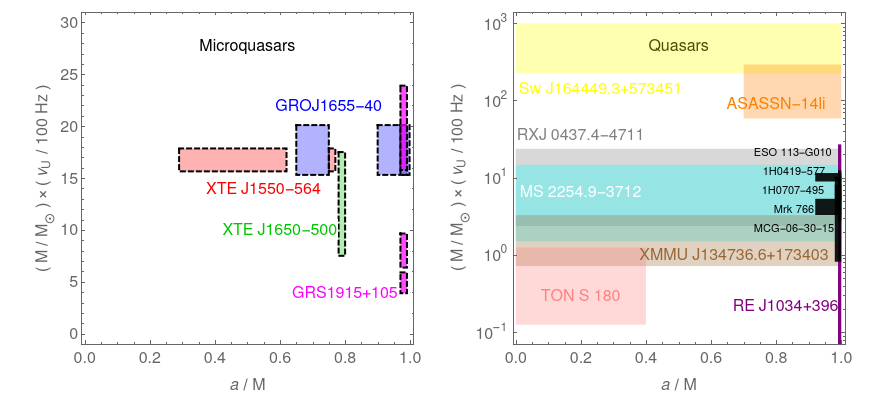}
\end{center}
\caption{The position of Quasars, and microquasars depending on the product of mass and frequencies as well as the dimensionless spin parameter $a$. The shaded regions represent the objects with mass estimates, often with large errors, but no spin estimates in the literature. The colored blocks specify those objects for which the spin is estimated, while the black regions indicate those objects for which only lower limit of spin is known.
\label{Quasars}}
\end{figure*}


\begin{figure*}
\flushleft
\includegraphics[width=0.32\hsize]{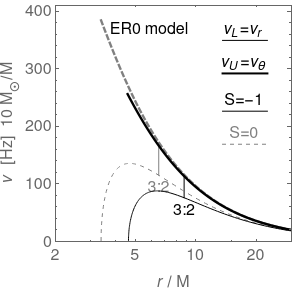}
\includegraphics[width=0.32\hsize]{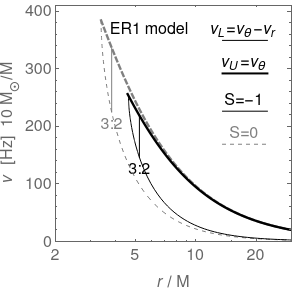}
\includegraphics[width=0.32\hsize]{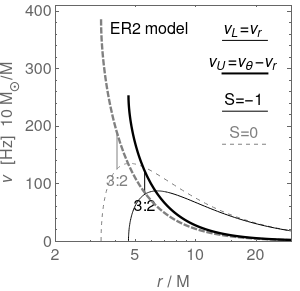}
\includegraphics[width=0.32\hsize]{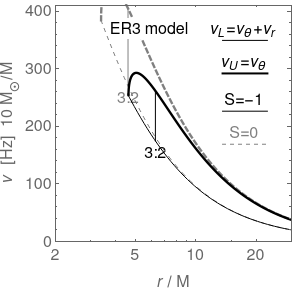}
\includegraphics[width=0.32\hsize]{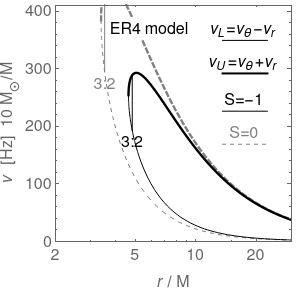}
\includegraphics[width=0.32\hsize]{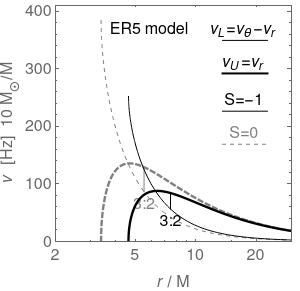}
\includegraphics[width=0.32\hsize]{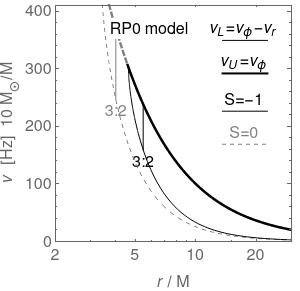}
\includegraphics[width=0.32\hsize]{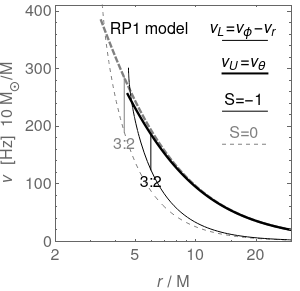}
\includegraphics[width=0.32\hsize]{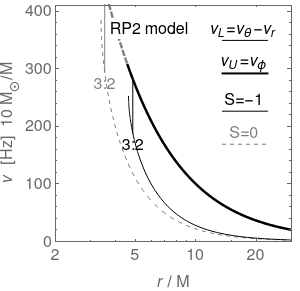}
\includegraphics[width=0.32\hsize]{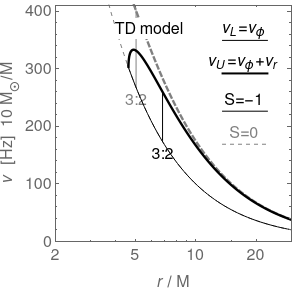}
\includegraphics[width=0.32\hsize]{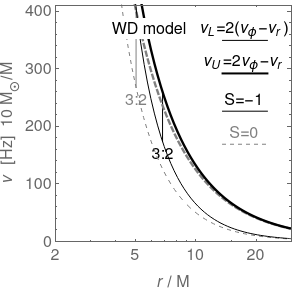}
\caption{Radial profiles of lower $\nu_{\rm L}(r)$ and upper $\nu_{\rm U}(r)$ frequencies for various HF QPOs models, see section \ref{models}. We compare the frequencies for neutral non-spinning particles ($S=0$, gray curves) with the neutral spinning particles ($S=-1$, black curves) orbiting Kerr BH. The position of $\nu_{\rm U}:\nu_{\rm L}=3:2$ resonance radii $r_{3:2}$ is also plotted. Black hole spin is taken to be $a=0.7$ in all cases.
\label{rprofile}
}
\end{figure*}

\section{Quasi-periodic oscillations models} \label{observations}

Spinning particle oscillations around circular orbits, studied in the previous section, suggest interesting astrophysical application, related to HF QPOs observed in selected quasars and microquasars. In the following subsection, we discuss the general constraint methods of BH parameters from QPOs on the predictions of the rotation and mass parameters of BHs. In the later subsections we describe the general technique useful for the QPO fittings and consider different QPO models, namely: epicyclic resonance (ER) model and its variants (ER1, ER2, ER3, ER4, ER5), relativistic precession (RP) model and its variants (RP1, RP2), tidal disruption (TD) model and warped disc (WD) model.


The twin peaks of the~HF~QPOs with upper $f_{\mathrm{U}}$ and lower $f_{\mathrm{L}}$ frequencies are sometimes observed in the~Fourier power spectra. In the microquasars and quasars the twin HF QPOs appear at the fixed frequencies that usually have nearly exact $3:2$ ratio \cite{McC-etal:2011:CLAQG:,Smi-Tan-Wag:2021:APJ:}. The observed high frequencies are close to the orbital frequency of the~marginally stable circular orbit representing the~inner edge of the accretion disks orbiting BHs; therefore, the~strong gravity effects are believed to be relevant for the explanation of HF~QPOs \cite{Tor-etal:2005:ASTRA:}.
The models of twin HF QPOs involving the orbital motion of matter around BH can be generally separated into four classes: the hot spot models (the relativistic precession model and its variations \cite{Ste-Vie:1999:PHYSRL:,Stu-Kot-Tor:2013:ASTRA:}, the tidal precession model \cite{Kos-etal:2009:ASTRA:}), resonance models \cite{Tor-etal:2005:ASTRA:,Stu-etal:2015:ACTA:} and disk oscillation (diskoseismic) models \cite{Rez-etal:2003:MNRAS:,Mon-Zan:2012:MNRAS:}. These models were applied to match the twin HF QPOs and the LF QPO for the microquasar GRO J1655-40 in \cite{Stu-Kol:2016:ASTRA:}. Of course, the models can be applied also for intermediate massive BHs \cite{Stu-Kol:2015:MNRAS:}.
Unfortunately, none of the models recently discussed in literature, based on the frequencies of the harmonic geodesic epicyclic motion, is able to explain the HF QPOs in quasars and all four microquasars simultaneously, assuming that their central attractor is a BH \cite{Tor-etal:2011:ASTRA:,Smi-Tan-Wag:2021:APJ:}. One of the most promising ways of explaining these QPO phenomena in quasars and microquasars could be charged particle oscillations around magnetized BHs \cite{Kol-Stu-Tur:2015:CLAQG:,Tur-Stu-Kol:2016:PHYSR4:,Kol-Tur-Stu:2017:EPJC:,Kol-Sha-Stu:2020:EPJC}.

\subsection{Tested HF QPOs models \label{models}}

The hot spot models assume radiating spots in thin accretion discs following nearly circular geodesic trajectories. In the standard RP model \cite{Ste-Vie-Mor:1999:ApJ:}, the upper of the twin frequencies is identified with the orbital (azimuthal) frequency, $\nu_{\rm U}=\nu_{\rm \mip}$, while the lower one is identified with the periastron precession frequency, $\nu_{\rm L}=\nu_{\rm \mip} - \nu_\mir$. The radial profile of the frequencies $\nu_U$ and $\nu_L$ of the RP model is presented in Fig. \ref{rprofile}.
The epicyclic resonance (ER) models \cite{Abr-Klu:2001:AA:,Tor-etal:2005:ASTRA:} consider a resonance of axisymmetric oscillation modes of accretion discs. Frequencies of the disc oscillations are related to the orbital and epicyclic frequencies of the circular geodesic motion. The radial profile of the frequencies $\nu_U$ and $\nu_L$ of the ER model is presented in Fig. \ref{rprofile}.
The tidal disruption TD model, where $\nu_{\rm U}=\nu_{\rm \mip} + \nu_\mir$ and $\nu_{\rm L}=\nu_{\rm \mip}$, could resemble to some degree the hot spot models as numerical simulations of disruption of inhomogeneities (e.g. asteroids for microquasars or stars for quasars) by the BH tidal forces demonstrate existence of an orbiting radiating core in the created ring-like structure \cite{Kos-etal:2009:ASTRA:}. The warped disc WD oscillation model of twin HF QPOs assumes non-axisymmetric oscillatory modes of a thin disc \cite{Kat:2008:PASJ:}.

\begin{figure*}
\flushleft
\includegraphics[width=0.32\hsize]{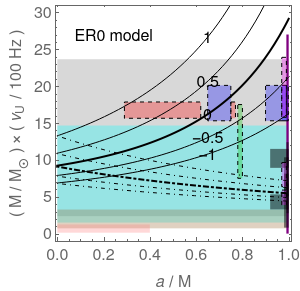}
\includegraphics[width=0.32\hsize]{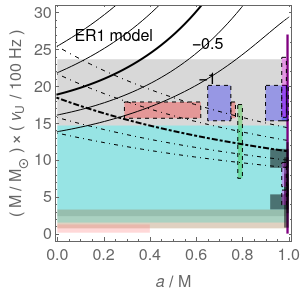}
\includegraphics[width=0.32\hsize]{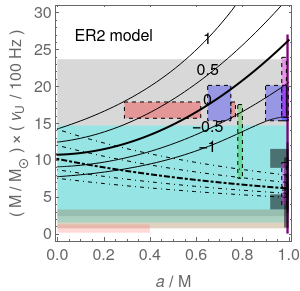}
\includegraphics[width=0.32\hsize]{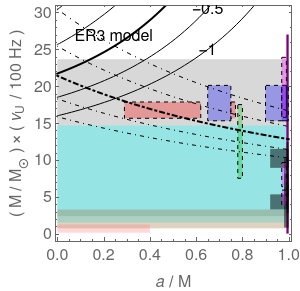}
\includegraphics[width=0.32\hsize]{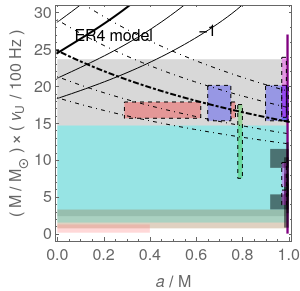}
\includegraphics[width=0.32\hsize]{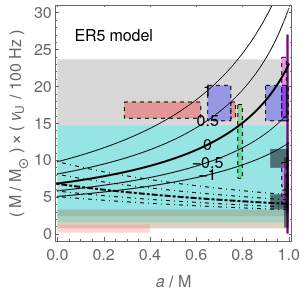}
\includegraphics[width=0.32\hsize]{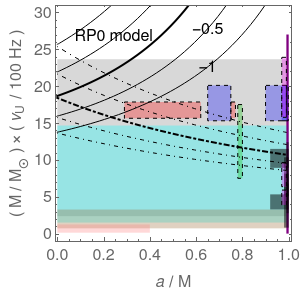}
\includegraphics[width=0.32\hsize]{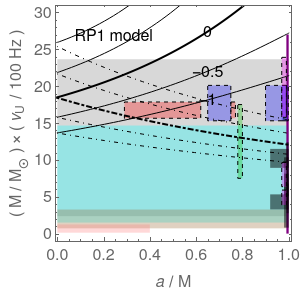}
\includegraphics[width=0.32\hsize]{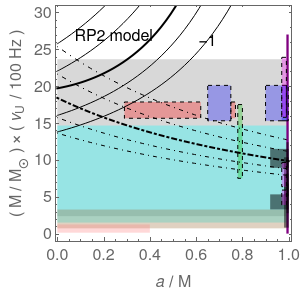}
\includegraphics[width=0.32\hsize]{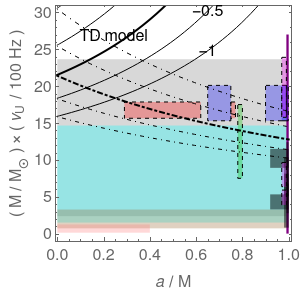}
\includegraphics[width=0.32\hsize]{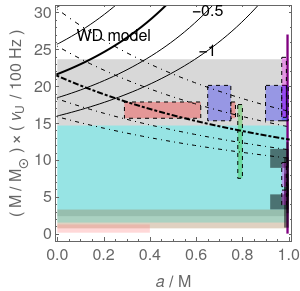}
\caption{Fitting of quasars and microquasars with different spinnig particle HF QPOs models. On the horizontal axis the BH spin $a/M$ is given, while on the vertical axis we give BH mass $M$ multiplied by observed upper HF QPOs frequency $\mu_{\rm U}$. Boxes represent the observed mass-frequency $M \cdot f$ and spin $a$ values for various quasar and microquasar sources (microquasar sources have dashed boundary), see table \ref{tab1}, and Fig.~\ref{Quasars}.
The curves represent the test particle frequencies at $r_{3:2}$ resonant radii $f_{\rm U} = \nu_{r}(r_{3:2},a,S)$ as function of BH spin $a$ for different values of spin parameter $S$. The solid curves are plotted for co-rotating spinning particles while dot-dashed curves for contra-rotating particles and the thick curves (solid and dashed) show the non-spinning fit ($S=0$). The numbers on the curves represent the value of spin parameter $S$. \label{fitsfig} }
\end{figure*}

 \subsection{Resonant radii and the fitting technique}

The HF QPOs come in pair of two peaks with upper $f_{\mathrm{U}}$ and lower $f_{\mathrm{L}}$ frequencies in the timing spectra. The frequency ratios $f_{\mathrm{U}}:f_{\mathrm{L}}$ are very close to the fraction $3:2$ indicating the existence of the resonances between two modes of oscillations. In case of geodesic QPO models, the observed frequencies are associated with linear combinations of the particle fundamental frequencies $\nu_{r}, \nu_{\theta}$ and $\nu_{\phi}$. In the case of spinning neutral particle orbiting Kerr BH, the upper and lower frequencies of HF QPOs are the functions of spin parameter $S$, rotation $a$, BH mass $M$, and the resonance position $r$,
\beq
 \nu_{\mathrm{U}} = \nu_{\mathrm{U}}(r,M,a,S), \quad \nu_{\mathrm{L}} = \nu_{\mathrm{L}}(r,M,a,S). \label{ffUL}
\eeq
It is worth to note that the frequencies $\nu_{\mathrm{U}}$ and $\nu_{\mathrm{L}}$ are inversely proportional to the mass $M$ of a BH, while the dependence of frequencies on the BH spin $a$ and orbiting particle spin $S$ is more complicated and hidden inside $\Omega_{r},\Omega_{\theta},\Omega_{\phi}$ functions, as given by Eq. (\ref{nu_rel}). In order to fit the frequencies observed in HF QPOs with the BH parameters, one needs first to calculate the so called resonant radii $r_{3:2}$
\beq
\nu_{\mathrm{U}}(r_{3:2}):\nu_{\mathrm{L}}(r_{3:2})=3:2 \label{rezrad}.
\eeq
Resonant radii $r_{3:2}$ in general case are given as the numerical solution of higher order polynomial in $r$, for given values of spin $a$. Since the Eq. (\ref{rezrad}) is independent of the BH mass explicitly, the resonant radius solution also has no explicit dependence on the BH mass and techniques introduced in \cite{Stu-Kot-Tor:2013:ASTRA:} can be used. Substituting the resonance radius into the Eq. (\ref{ffUL}), we get the frequency $\nu_{\mathrm{U}}$ in terms of the BH mass, rotation $a$ and particle spin $S$.
Now we can compare calculated frequency $\nu_{\mathrm{U}}$ with observed HF QPOs frequency $f_{\mathrm{U}}$, see Tab.~\ref{tab1}. Calculated upper frequency at resonant radii $r_{3:2}$ as function of BH spin $\nu_{\mathrm{U}}(a)$ has been used to fit observed BH spin and mass data in Fig.~\ref{fitsfig} for various QPO models. For particle internal spin $S$ we use values $(-1,-0.5,0,0.5,1)$.

\section{Astrophysical relevance of test particle spin} \label{spin}

The system we are considering in this paper is a compact spinning body of mass $\mu$ orbiting a large body of mass $M$, which taken to be a solar or supermassive Kerr BH, and satisfy the condition $\mu \ll M$. The spin parameter is measured in $\mu M$ units, namely $S/(\mu M)$, hence the test particle spin $S$ will a dimensionless quantity. As for the spin of a particle with mass $\mu$, we usually expect the condition $S \leq O(\mu^{2})$, therefore for the present case, we have
\beq
S = \frac{S}{\mu M} = \frac{S}{\mu^2} \frac{\mu}{M} \leq O \left (\frac{\mu}{M} \right) \ll 1.
\eeq
Thus the physically realistic values of the spin parameter must satisfy $S \ll 1$ for the
compact objects (neutron stars, BHs, and white dwarfs). The most models of neutron stars have the spin bound $S \leq 0.6 ~ \mu/M$. The realistic values of spin parameter $S$ for LISA sources fall in the range $10^{-4}-10^{-7}$. For a central BH having mass $M = 10^{6} M_{\odot}$, the spin parameter bound is $S \leq 9 \times 10^{-6}$ (corresponding to a white dwarf with $\mu = 0.5 M_{\odot}$) \cite{Hartl:2003:PRD:}. Furthermore, it is worthwhile to mention that for $S>1$, the Papapetrou equations are not physically realistic, since they are derived in the limit of spinning test particles, which must satisfy $\mu \ll M$. Thus one cannot draw reliable results about the behavior of astrophysical systems for $S > 1$.

The oscillating test particle could represent whirl or spinning hot spot in plasma moving around central BH. It could also represent spinning asteroid for microquasars or rotating star, neutron star in the case of quasars. Obviously the maximal spin $S=\pm1$ used in QPOs fits Fig.~\ref{fitsfig} is upper bound for any realistic situation and the spin of any astrophysicaly relevant object must be lower.

\section{Discussion and Conclusions} \label{kecy}

In this article we have studied classical spinning test particle motion around rotating Kerr BH and applied spinning particle fundamental frequencies to observe the QPOs in quasars as well as microquasars.

There are four possible situations of spinning test particle motion depending on the direction of the BH rotation $a$ and particle spin $S$. The co-rotating particles of ISCO shift towards the BH horizon with increase of the BH rotation $a$, while counter-rotating particles shift away from the BH horizon. Furthermore, the position of ISCO is situated close to the horizon when the particle spin is aligned to the z-axis as compared to the case when the spin is anti-aligned.

We have explored the particle fundamental frequencies of small harmonic oscillations around circular orbit. The behavior of frequencies lowering down or increasing up depends on the direction of the BH rotation parameter and test particle spin. Both parameters $a$ and $S$ contribute to increase the fundamental frequencies, however, higher frequencies can be observed when the spin is perpendicular to the equatorial plane while lower frequencies in the opposite case. It is also noticed that co-rotating particles have higher frequencies in comparison with the contra-rotating particles.

The radial profiles of lower $\nu_{\rm L}(r)$ and upper $\nu_{\rm U}(r)$ frequencies for various HF QPOs models (RP, RP1, ER0, ER1, ER2, ER3, ER4, ER5, TD and WD) has been examined and fitted the observed QPOs data by spinning particle fundamental frequencies. The spinning particle QPOs models can fit the data for all four microquasars (GRS~1915+105, GRO~1655-40, XTE~1550-564, XTE~J1650-500) and also some of quasar sources. Unfortunately there is no spinning particle QPOs model which can fit all quasar sources: the sources (Sw~J164449.3+573451, ASASSN-14li) with too high frequency mass scale ($f~\times~M>100$) and the sources (TON~S~180, XMMU~J134736.6+173403) with too low frequency mass scale ($f~\times~M<5$) can not be fitted.

All test particle spin effects are relevant for high spins $|S|\sim~1$ only. Test particle spin for astrophysically realistic object is low $S\leq1$, only in the case of BH-BH mergers where large spins are relevant only can assume significant shifts in spinning particle frequencies \cite{Sko-Ger:2021:NN:}.

\section*{Acknowledgments}

The authors M.K. and Z.S. would like to express their acknowledgments for the Research Centre for Theoretical Physics and Astrophysics, Institute of Physics, Silesian University in Opava, Z.S. acknowledge the Czech Science Foundation Grant No. 19-03950S.



\def\prc{Phys. Rev. C}
\def\pre{Phys. Rev. E}
\def\prd{Phys. Rev. D}
\def\prl{Physical Review Letters}
\def\jcap{Journal of Cosmology and Astroparticle Physics}
\def\apss{Astrophysics and Space Science}
\def\mnras{Monthly Notices of the Royal Astronomical Society}
\def\apj{The Astrophysical Journal}
\def\aap{Astronomy and Astrophysics}
\def\actaa{Acta Astronomica}
\def\pasj{Publications of the Astronomical Society of Japan}
\def\apjl{Astrophysical Journal Letters}
\def\pasa{Publications Astronomical Society of Australia}
\def\nat{Nature}
\def\physrep{Physics Reports}
\def\araa{Annual Review of Astronomy and Astrophysics}
\def\apjs{The Astrophysical Journal Supplement}
\def\aapr{The Astronomy and Astrophysics Review}
\def\procspie{Proceedings of the SPIE}


\end{document}